%% file: studivz.tex
\documentclass[authoryear]{elsarticle}

\usepackage{amsmath} \usepackage{dcolumn} 
\usepackage{graphicx} \usepackage{ifpdf}
\usepackage{url} \usepackage{xspace}

\newcolumntype{d}[1]{D{.}{.}{#1}} \newcolumntype{.}{D{.}{.}{-1}}

\ifpdf
\usepackage{pdfsync}
\else
\fi

\usepackage{hyperref} 
\usepackage{cleveref}

\include{studivzmacros}

\journal{Social Networks}
\crefname{equation}{Eq.}{Eqs.}  \crefname{figure}{Fig.}{Figs.}
\crefname{section}{Section}{Sections}
\crefname{subsection}{Section}{Sections}
\crefname{subsubsection}{Section}{Sections}
\crefname{table}{Table}{Tables}

\graphicspath{{pdffigs/}{epsfigs/}{mpsfigs/}}

\begin{document}

\begin{frontmatter}

\title{Investigating an online social network using spatial interaction models}

\author[1]{Conrad Lee\corref{cor}\fnref{altadd}}
\author[2]{Thomas Scherngell}
\author[2]{Michael J. Barber}

\cortext[cor]{Corresponding author}
\fntext[altadd]{\textit{E-mail}: conradlee@gmail.com}
\address[1]{Clique Research Cluster, University College Dublin, Ireland}
\address[2]{AIT Austrian Institute of Technology GmbH, Foresight and Policy Development Department, Vienna, Austria}

\begin{abstract}
  Using a spatial interaction modeling approach, we investigate
  the German collegiate social network
  site \studivz. We focus on identifying factors that foster strong 
  inter-institutional linkages, 
  testing whether the acquaintanceship rate between
  institutions of higher education is related to various geographic
  and institutional attributes. We find that acquaintanceship is most
 significantly related to geographic separation: measuring distance
 with automobile travel time,  acquaintanceship drops by 91\% for
 each additional 100 minutes. Institution type and
  the former East-West German divide are also related to this
  rate with statistical significance.
\end{abstract}

\begin{keyword}
online social networks \sep spatial interaction modeling \sep
generalized linear models \sep separation measures \sep StudiVZ
\end{keyword}
\end{frontmatter}


\section{Introduction} \label{sec:introduction} Recent research of
social network data indicates that the strength of flow between cities
can be approximated by a gravity law. That is, the flow between a pair
of cities---which may represent people, communication, goods, or
money---is proportional to the product of their sizes divided by the
square of their distance \citep{lambiotte-2008,Krings2009}.  However,
flows between cities depend upon more than their sizes and the
distance that separates them; they also depend upon cultural factors
such as language, region, or economic similarity. For example, the
work of \citet{Blondel2008} suggests that the Belgian mobile
communication network splits rather cleanly into two large-scale
communities: French and Dutch speakers.  Using the network of flows of
currency between cities, \citet{Brockmann2010} found that flows do not
continuously decrease as a function of distance, but rather are
affected by sharp cultural boundaries embedded in geographic space.

Given a social network that contains data on both geographic location
and various other node attributes such as language, it is natural to
hypothesize that some node attribute is related to geographic flow. In
this paper, we argue that spatial interaction models are useful for
statistically testing such hypotheses.  Spatial interaction modeling
focuses on estimating the variation of flows between locations in
geographic space, such as regions or cities \citep[see, for
example,][]{SenSmi:1995} While spatial interaction modeling is widely
used in the fields of economics and economic geography to understand
and predict interaction behavior in a system of spatial units (for an
overview, see \citet{FotOKe:1989}), it has only rarely been applied to
social network data \citep{SchBar:2009}.

We use spatial interaction models to investigate
data from the social network site (SNS) \studivz, a collegiate online
social network popular in German-speaking countries.  In particular,
we consider acquaintanceships between students at the University of
Bielefeld (\uniB) and students at other institutions. We determine how
various geographic and institutional attributes---including
institution type and region---affect the number of acquaintanceships
between these institutions.

We use these attributes to define several measures of separation
between institutions, with their importance assessed by comparing to
the data.  We consider two similar models, the first with fewer and
rougher separation measures, and the second with more and finer
separation measures.  For both models, we find that the most
pronounced separation measure is geographic distance, with institution type (\eg, university, technical college, art
school) also playing an important and significant role.  Regional
factors play a less pronounced, yet statistically significant, role.

In \cref{sec:studivz}, we give an overview of \studivz and describe
how we collected the data on its linkage structure. We begin
\cref{sec:spatialinteraction} by briefly introducing spatial
interaction modeling in general; we then explain the details of the two
particular models that we propose. Estimation results
for the models are presented in \cref{sec:results}. We summarize and
discuss our results in \cref{sec:conclusion}.

\section{\studivz} \label{sec:studivz}

\subsection{Overview}

\studivz is a social network site for German post-secondary students
that was created in October 2005. As reported by \citet{bakst-2006},
\studivz was inspired by and intentionally imitated the leading
U.S. social networking service for students, Facebook, and the site
therefore shares many features with Facebook. Within months of its
creation, \studivz became the dominant social networking service for
students in Germany, Austria, and regions of Switzerland.

Social network sites have been defined by \citet{boyd-2007} as web
services that allow individuals to (1) construct a public or
semi-public profile, (2) articulate a list of other users with whom
they are connected, and (3) view and traverse connections made by
others.  In addition to these basic capabilities, important functions
of \studivz and most other SNSs include semi-public photo-sharing,
``walls'' (\ie, message boards), and status updates---all linked to
individuals' profile pages. Users of SNSs are often highly engaged;
for example, at the time we collected data, \studivz users averaged
about one visit per user per day, and about 30 page impressions per
visit \citep{studivz-2008}.

While one might expect much of the interaction between users of these
sites to take place between people who have met only on the internet
(as is often the case for message boards and chat rooms),
\citeauthor{boyd-2007} note that the relationships articulated in
SNSs tend to have an offline origin. This claim has been supported by
the findings of \citet{mayer-2008}, who report that only $0.4\%$ of
the Facebook friendships they studied appear have originated as
``merely online friendships,'' as well as other empirical studies
\citep{haythornthwaite-2005,lampe-2006,ellison-2007}. This offline
basis suggests that conclusions based on SNS data are more likely to
be generalizable to everyday, personal social interaction than data
which comes from other online services such as chat rooms or message
boards, where social interaction is often anonymous.

Despite the correspondence between online and offline social contact,
it is difficult to interpret the meaning of SNS relationships, labeled
in \studivz as ``friendships.'' In the \studivz data that we analyze
in this paper, over half of users listed more than 48 ``friends,'' and
over a quarter of users had more than 86 friends. In a recent dataset
from the more mature Facebook, where users have had more time to
accumulate friends, over half of users listed more than 100 friends
\citep{lewis-2008}. Such large degrees indicate that many of these
so-called friendships are not close, active friendships, but rather
latent friendships or acquaintanceships, such as old high-school
friends or colleagues from previous jobs.  Nevertheless, for the rest
of the paper, all of these ties will be referred to as friendships.

Within the \studivz system, users associate themselves with the
institutions they attend.  As we collected data in January 2008,
\studivz reported 4.5 million members \citep{studivz-2008}. If one
considers that a total of 2.47 million students were enrolled at
higher-learning institutions in Germany, Austria, and Switzerland in
the academic year 2007/2008,\footnote{In the academic year 2007/2008,
  Germany reported 1.97 million post-secondary students
  \citep{hsk-2008}, Austria reported 272,003
  \citep{austro-bildung-2009}, and Switzerland reported 225,862
  thousand \citep{schweiz-2009}.}  it appears that not only had
\studivz largely saturated its target audience of students, but also
that the site had attracted users from other audiences, perhaps former
students, exchange students, and young people who were not students.
These users presumably associated themselves with institutions that
they did not attend.

\subsection{Data Collection}

We retrieved data in late January, 2008. The data collected that is
relevant to this paper is the friendship data, which at the time of
collection was publicly visible for all profiles. As mentioned above,
a ``friendship'' in \studivz is the primary indicator of relation, and
is formed when one user requests another user to confirm him as a
friend. The information concerning the direction of the request (i.e,
who requested whom) is not public, and therefore we treat all friendships as undirected.

We collected the friendship data using \newterm{snowball sampling},
such that we were left with all friendship data of \uniB users in the
giant connected component of \studivz's \uniB subgraph.  This data set
included 29,192 users. The users we identified at \uniB listed over
1.4 million friendships. Around 350,000 of these friendships are among
the 29,192 \uniB users; we refer to these as \newterm{internal
  friendships}. Approximately 1.05 million friendships were between
\uniB students and some 367,000 students associated with other
higher-learning institutions; we call these \newterm{external
  friendships}.  Because at the time of collection \studivz enjoyed
such widespread popularity among German college students, we contend
that a considerable proportion of all acquaintanceships between the
students of the University of Bielefeld and students at other
institutions is included in the external friendships.

Here we examine only the external friendships aggregated at the level
of institutions of higher education (henceforth labeled as simply
``institutions'').  More specifically, a weight was determined for the
connection between each of \numinst German institutions and \uniB.
The weight between \uniB and some other institution
corresponds to the number of friendships between all \studivz users in
\uniB and all \studivz users at that institution.

We exclude from consideration the four nearest institutions, each with
a travel time of under thirty minutes to \uniB, because these
institutions have special relationships with \uniB (\eg, shared
dormitories, libraries, and presumably other, unknown arrangements).
These relationships are not accounted for in our model, leading it to
drastically under-predict the rate of friendship with these four institutions. 
Other institutions that are nearby, but outside this thirty-minute limit, do not 
show a marked mismatch with the model predictions.
Additionally, we
consider only institutions in Germany, for which information on
addresses, institution type, and enrollment size is consistently
available from the website of the German Rectors'
Conference.\footnote{\url{http://www.hochschulkompass.de/hochschulen/download.html}}

\section{Spatial Interaction Model
  Definition} \label{sec:spatialinteraction}

To investigate the \studivz network, we adopt a \newterm{spatial
  interaction modeling} perspective. Spatial interaction modeling
focuses on estimating the variation of flows between locations in
geographic space, such as regions or cities \citep[see, for
example,][]{SenSmi:1995}.  Said flows, known as \newterm{spatial
  interactions}, can be of many forms, including people, goods, money,
or knowledge.  Spatial interaction modeling is widely used in the
fields of economics and economic geography to understand and predict
interaction behavior in a system of spatial units; for an overview,
see \citet{FotOKe:1989}.  Here, we will investigate flows of social
interactions, in the form of online friendships.

The specific technique we will use is that of \newterm{generalized
  linear models}. The usual approach is to assume that the spatial
interaction \( \floworigindestination{i}{j} \) between an origin
location \( i \) and a destination location \( j \) has a product form
given by
\begin{equation}
  \flowmodelorigindestination{i}{j} = \originfunction{i} \destinationfunction{j} \separationfunction{i}{j}
  \mathcomma
  \label{eq:origindestinationmodel}
\end{equation}
where \( \originfunction{i} \) is the \newterm{origin function}, \(
\destinationfunction{j} \) is the \newterm{destination function}, and
\( \separationfunction{i}{j} \) is the \newterm{separation function}.
As \studivz friendships are inherently symmetric, we can without loss
of generality assume that \uniB is always the destination, reducing
the model to a one-dimensional version given by
\begin{equation}
  \flowmodel{i} = \destconst \originfunction{i} \sepfunc{i}
  \mathperiod
  \label{eq:spatialinteractionmodel}
\end{equation}
In \cref{eq:spatialinteractionmodel}, \( \flowmodel{i} \) and \(
\sepfunc{i} \) are the source-only analogs of \(
\flowmodelorigindestination{i}{j} \) and \( \separationfunction{i}{j}
\), respectively. The destination function \( \destinationfunction{j}
\) has been replaced by a constant term \( \destconst \), reflecting
the contribution from \uniB. As \( \originfunction{i} \) and \(
\sepfunc{i} \) both have the same dependence on \( i \), the two terms
could be merged into one, but we will treat them separately to
maintain the distinction between properties of just the origin
institution and properties dependent upon the separation between the
origin and destination (\ie, \uniB) institutions.

We take the origin function \( \originfunction{i} \) to be
\begin{equation}
  \originfunction{i} = \size{i}^{\originparameter}
  \mathcomma
  \label{eq:originfunction}
\end{equation}
where \( \size{i} \) is the number of students enrolled at institution
\( i \) during the 2007/2008 academic year, accounting for differences
in the sizes of the institutions; \( \originparameter \) is a
parameter to be determined.  We take the separation function \(
\sepfunc{i} \) to include \( \numsepmeasures \) measures of
separation, with a general form of
\begin{equation}
  \sepfunc{i} = \exp \largeparens{\sum_{k=1}^{\numsepmeasures} \distanceparameter{k} \distance{k}{i}}
  \mathperiod
  \label{eq:separationfunction}
\end{equation}
The parameters \( \distanceparameter{k} \) weigh various separation
measures \( \distance{k}{i}\) against one another. For notational
convenience, we similarly rewrite the destination constant as an
exponential, \( \destconst = \exp{\logdestconst} \).

With the origin and separation functions chosen as in
\cref{eq:originfunction,eq:separationfunction}, the spatial
interaction model in \cref{eq:spatialinteractionmodel} is a
generalized linear model. A common approach for determining the model
parameters is standard OLS estimation \citep{BerWes:1997}.  However,
as the \( \flowemp{i} \) are non-negative integers, OLS estimation is
inappropriate, being equivalent to assuming the residuals \(
\flowmodel{i} - \flowproductmodel{i} \) for
\cref{eq:spatialinteractionmodel} are normally distributed.  In the
present case, a discrete data generating process would then be
approximated by a misrepresentative continuous process.  Instead, we
assume a negative binomial model specification
\begin{equation}
  P \largeparens{\flowemp{i}} = 
  \frac{ \gammafunc{\flowemp{i} + \inversedispersionparameter} }{ \gammafunc{\flowemp{i} + 1}\gammafunc{\inversedispersionparameter} }
  \largeparens{\frac{\inversedispersionparameter}{\flowproductmodel{i}+\inversedispersionparameter}}^{\inversedispersionparameter}
  \largeparens{\frac{\flowproductmodel{i}}{\flowproductmodel{i}+\inversedispersionparameter}}^{\flowemp{i}}
  \mathcomma
  \label{eq:negbinspec}
\end{equation}
where \( \gammafunc{\cdot} \) denotes the Gamma function and \(
\dispersionparameter \) is a dispersion parameter; in the limit 
as \( \dispersionparameter \approaches 0 \), \cref{eq:negbinspec} reduces 
to a Poisson model specification.  We estimate model
parameters with maximum likelihood procedures using
Newton-Raphson \footnote{For parameter estimation, we use the open source statistical software R (version R-2.12.0).}. For more detail on the derivation and properties of
negative binomial models like those in \cref{eq:negbinspec}, see
\citet{CamTri:1998}.

The separation function constitutes the core of a spatial interaction
model, and is of central importance in the context of the research
questions of the current study. We use a multivariate exponential
separation function \citep{SenSmi:1995}, as defined in
\cref{eq:separationfunction}, that provides a flexible framework for
representing different kinds of separation.  In the current paper, we
focus on \( \numsepmeasures = 10 \) separation measures.

We distinguish between geographic separation effects, separation
effects related to the federal states of Germany (which will
henceforth be referred to simply as \textit{states}), and
institutional separation effects. Geographic separation effects are
captured by \( \logtraveltimesepvar{i} \), the logarithm of the
automobile travel time\footnote{We selected the travel time as
  corresponding to the geographic separation as experienced for
  personal interactions.  Other measures of geographic separation
  include the great circle distance or driving distance between the
  institutions; these are strongly correlated with the travel time.
  To calculate the travel time between the University of Bielefeld and
  another institution, we queried Google Maps
  (\url{http://maps.google.de/}) for the drive time between the
  University of Bielefeld's address and the other institution's
  address using the default routing settings.} in minutes between
\uniB and the other institutions. We augment this with \(
\samebundeslanddummy{i} \), a dummy variable that measures separation
effects related to the states of Germany. It is defined as a binary
variable set to one if institution \( i \) is located in North
Rhine-Westphalia, the same state as \uniB, and zero otherwise.

Further, we take into account separation effects related to the type
of the institutions. The dummy variable \( \samecategorydummy{i} \)
accounts for the type of institution, with its value set to one if one
institution \( i \) is the same type of institution as \uniB, and zero
otherwise. Here we use information on the organization of the German
post-secondary education system, which contains several types of
institutions. Our dataset includes general universities\footnote{We
  define ``general universities'' to be public institutions that have
  the right to grant doctoral degrees and which offer a full range of
  faculties.  For example, we categorize technical universities as
  general universities, but not dedicated medical schools, applied
  technical colleges, art schools, or those institutions which are
  private, religious, or run by the military.  \uniB itself is a
  general university and is located in former West Germany.};
\foreign{Fachhochschulen} (officially translated as ``universities of
applied sciences'', and roughly described as technical colleges or
trade schools---these will be referred to as ``applied
universities''); religious institutions; art schools, including music
schools; teacher colleges; and private institutions.

With this information we further refine \( \samecategorydummy{i} \)
into separate dummy variables for each institution type, more
precisely reflecting how each type of institution affects interaction
probability. Thus, we set a value of one for \( \fhdummy{i} \) if
institution \( i \) is an applied university, for \( \kirchendummy{i}
\) if institution \( i \) is a religious, for \( \kunstdummy{i} \) if
institution \( i \) is an art school, for \( \paeddummy{i} \) if
institution \( i \) is a teacher college, for \( \privatdummy{i} \) if
institution \( i \) is a private institution, and for \( \unidummy{i}
\) if institution \( i \) is a general university.

Due to the enormous political and institutional changes during the
reunification of the Federal Republic of Germany and the German
Democratic Republic in 1990, we considered a variable that reflects
predominant barriers between these major blocks. Thus, we introduce \(
\ostendummy{i} \), a dummy variable that is set to one if institution
\( i \) is located in the former German Democratic Republic, and zero
otherwise.

\section{Estimation Results} \label{sec:results}

In this section, we discuss the estimation results of the negative
binomial spatial interaction model, as defined by
\cref{eq:negbinspec}.  The dependent variable is the observed number
of friendships between institution \( i \) and the University of
Bielefeld. The independent variables are origin and separation
measures as defined in \cref{sec:spatialinteraction}. We consider a
basic model version including separation measures for the logarithmic
geographic distance \( \logtraveltimesepvar{i} \), the same state \(
\samebundeslanddummy{i} \), and the same category \(
\samecategorydummy{i} \). In an extended model version, we replace \(
\samecategorydummy{i} \) with specific dummy variables for each
category (\( \distance{k}{i} \) for \( k=4, 5, \ldots, 9 \)).  The
extended model serves to check robustness of the other model
parameters \( \logtraveltimesepvar{i} \) and \(
\samebundeslanddummy{i} \), while also providing additional
information on specific effects related to different institution
types.

In \cref{tbl:spatialinteractionmodel}, we present the sample estimates
of the spatial interaction model parameters, along with tests of
statistical significance \citep{Gre:2003}. The second and third
columns contain modeling results for the basic and extended models,
respectively.  There are \numinst observations. The estimate of the
dispersion parameter \( \dispersionparameter \) is 0.51 in the basic
model version and 0.45 in the extended model version; the estimates
are statistically significant. Thus, we conclude that unobserved
heterogeneity, not captured by the covariates, leads to overdispersion
\citep{CamTri:1998} and that the negative binomial specification of
\cref{eq:negbinspec} is a more appropriate choice than would be, \eg,
a simpler Poisson specification.

\begin{table}
  \begin{tabular}{l..}
    \hline\hline
    & \multicolumn{2}{c}{Model Version} \\ 
    \cline{2-3}
    & \columnhead{c}{Basic} & \columnhead{c}{Extended} \\ 
    \hline
     Destination constant \( \logdestconst \) & 9.37(0.49)\esignificant & 9.59(0.55)\esignificant  \\
     Origin exponent \( \originparameter \) & 0.76(0.03)\esignificant & 0.71(0.05)\esignificant  \\
     Log geographic distance \( \logtraveltimeparameter \) & -1.45(0.09)\esignificant & -1.43(0.08)\esignificant  \\
     Same state \( \samebundeslandparameter \) & -0.32(0.16)\significant & -0.32(0.15)\significant  \\
     Same category \( \samecategoryparameter \) & -0.58(0.12)\esignificant & \emptycell  \\
     Applied University \( \fhparameter \) & \emptycell & -0.56(0.16)\esignificant  \\
     Religious Institution\( \kirchenparameter \) & \emptycell & -0.98(0.24)\esignificant  \\
     Art School \( \kunstparameter \) & \emptycell & -0.73(0.21)\esignificant  \\
     Teacher College \( \paedparameter \) & \emptycell & -0.26(0.31)  \\
     Private \( \privatparameter \) & \emptycell & -0.28(0.21)  \\
     General University \( \uniparameter \) & \emptycell & 0.20(0.17)  \\
     Former GDR \( \ostenparameter \) & \emptycell & -0.27(0.11)\significant  \\
     Dispersion parameter \( \dispersionparameter \) & 0.51(0.04)\esignificant & 0.45(0.04)\esignificant  \\
    \hline
    \multicolumn{1}{l}{McFadden's Adjusted \( R^2 \)} & 0.115 & 0.139 \\
    \multicolumn{1}{l}{Cragg-Uhler (Nagelkerke) \( R^2 \)} & 0.846 & 0.900 \\
    \hline
    \multicolumn{1}{l}{Log-likelihood} & -2106.52 & -2085.83 \\
    \multicolumn{1}{l}{Likelihood ratio \( \chisq{} \)} & 659.99 & 701.37 \\
    \multicolumn{1}{l}{Akaike Information Criterion} & 14.998 & 13.802 \\
    \multicolumn{1}{l}{Bayesian Information Criterion} & 2609.460 & 2502.287 \\
    \hline\hline
    \multicolumn{3}{l}{\( \significant p < 0.05\) \quad \( \vsignificant p < 0.01 \) \quad \( \esignificant p < 0.001 \)}\\
  \end{tabular}
  \caption{Parameter estimates for the spatial interaction models. Fit
    parameters for the spatial interaction models are based on
    \numinst institutions.  Parenthetical values show the 
    standard errors. 
    Asterisks indicate the statistical
    significance of the parameter fits. Performance statistics 
    indicate that the extended model better explains the data.}
  \label{tbl:spatialinteractionmodel}
\end{table}

The model diagnostics underpin the statistical significance of the
models.  The likelihood ratio test is significant for both model
versions. To compare the two models, we draw on the Bayesian
Information Criterion (BIC), which is widely used for comparing
model fit of non-nested models \citep[see, for instance, ][]{Raf:1995}.
It can be seen that the extended model version fits better to the
data. This is also reflected by McFadden's \( R^2 \) and 
Nagelkerke's \( R^2 \) in terms of the explained variance 
by the covariates 
\citep[for a formal description of the model diagnostics used, see][]{LonFre:2001}.

Turning our attention first to the results of the basic model in the
second column, we note that increasing travel time between the
University of Bielefeld and other institutions \( i \) has a
significant negative effect on the number of friendships.  This result
indicates that \studivz is not a virtual world in which geographic
location is unimportant.  The parameter estimate of \(
\logtraveltimeparameter = -1.45 \) in the basic model indicates that
for each additional 100 minutes of travel time to the University of
Bielefeld, the mean friendship probability decreases by 91\%.
The parameter \( \samebundeslandparameter \) is estimated to be -0.3
and is statistically significant, indicating that the rate of
acquaintanceship decreases when institution \( i \) is located in a
different state than the University of Bielefeld.  However, more
important than whether \( i \) is located in the same state as the
University of Bielefeld is whether \( i \) is in the same
institutional category. The parameter \( \samecategoryparameter \) is
estimated to be -0.7, and reveals an interesting tendency: students at
the University of Bielefeld, which is a general university, tend to
have a higher rate of acquaintanceship with students from other
general universities than with students attending other types of
institutions.

The parameter estimates of the extended model, displayed in the third
column, include more detailed information about how region and
institution type affect the rate of acquaintanceship.  While the
estimate for the same state remains unchanged at -0.3, our estimate
for the \( \ostenparameter \) parameter, also -0.3, indicates that
those \( i \) located in former East Germany tend to have a lower
rates of acquaintanceship than those \( i \) located in former West
Germany.  Furthermore, the more specific treatment of institution type
in the extended model reveals that \uniB has especially low
acquaintanceship rates with religious and art-focused institutions,
while the negative effects associated with applied technical colleges,
private institutions, and teacher colleges are less pronounced.

In the extended model version, the parameter estimate of log travel
time \( \logtraveltimeparameter \) slightly decreases to a value of \(
-1.43 \), indicating that the estimated parameter is quite robust. The
slight decrease might suggest that travel time is to some extent a
proxy for other unobserved characteristics in the basic model that are
better reflected by additional independent variables in the extended
model.

\section{Conclusion} \label{sec:conclusion}

In this paper, we have used spatial interaction models to investigate  
\studivz friendships aggregated at the level of institutions of higher education.
We have focused identifying institutional attributes that are
related to the strength of ties between institutions.   
Our results indicate that geographic
distance is the most important separation variable: for each
additional 100 minutes of drive time separating \uniB from some other
institution, the mean acquaintanceship rate with that institution
drops by approximately 90\%.  \studivz is thus not a virtual world in
which physical distance is insignificant.  This finding is in
agreement with a previous study of LiveJournal, a blogging service
with SNS features \citep{nowell-2005}.

Additionally, we found that students at \uniB are most likely to form
relationships with students at ``peer institutions,'' \ie, students at
other universities and technical universities, and that they are less
likely to form acquaintanceships with students at teacher colleges,
applied technical colleges, art schools, and religious institutions.
Acquaintanceship is also (albeit less prominently) affected by
regional characteristics: students at \uniB are more often acquainted
with students in their own federal state (North Rhine-Westphalia), and
to the same extent they more frequently are friends with students from
former West Germany than students from former East Germany.

One important line of future work is to confirm these results with
network data that is more complete. Our results are based on only
those edges that have one end connected to a student at \uniB, and they are therefore skewed to reflect the preferences of students at
that institution. With complete network data one could test whether
these preferences closely resemble those of German students in
general.

More broadly, the spatial interaction modeling approach taken in this paper
could be applied to other networks and with other separation measures. 
Depending on the data available, models of greater sophistication---either 
conceptually or methodologically---may become relevant. For example, spatial
interaction models may include production constraints or attraction 
constraints \citep{SenSmi:1995}, and spatial autocorrelation may be taken 
into account using spatial filtering techniques \citep{FisGri:2008}.

\acknowledgments
This work has been supported by the European FP6-NEST-Adventure
Programme (contract number 028875); by the FWF, the Austrian Science
Fund (project number P2154); by the Portuguese FCT and POCI 2010 
(project MAT/58321/2004)
with participation of FEDER; by the Science Foundation Ireland
under grant 08/SRC/I1407, Clique: Graph and Network Analysis Cluster;
and by a US-German Fulbright Grant.


\bibliographystyle{plainnat}

\bibliography{mjbrefs,clrefs}

\end{document}

%% file: studivzmacros.tex
\newcommand{\ie}{i.e.}
 
\newcommand{\eg}{e.g.}

\newcommand{\uniB}{UniBi\xspace}
\newcommand{\studivz}{StudiVZ\xspace}

\newcommand{\foreign}{\textit}
\newcommand{\newterm}{}

\newcommand{\columnhead}[2]{\multicolumn{1}{#1}{#2}}
\newcommand{\emptycell}{\multicolumn{1}{c}{---}}

\newcommand{\acknowledgments}
	{\section*{Acknowledgments}\label{sec:acknowledgments}} 

\newcommand{\numinst}{304\xspace}

\newcommand{\largeparens}[1]{\left ( #1 \right )}

\newcommand{\approaches}{\rightarrow}

\newcommand{\gammafunc}[1]{\Gamma\largeparens{#1}}
\newcommand{\chisq}[1]{\chi_{#1}^{2}}

\newcommand{\mathpuncspace}{\quad}
\newcommand{\mathcomma}{\mathpuncspace,}
\newcommand{\mathperiod}{\mathpuncspace.}

\newcommand{\size}[1]{a_{#1}}

\newcommand{\flowsymbol}{F}
\newcommand{\empiricalflowsymbol}{f}
\newcommand{\floworigindestination}[2]{\flowsymbol_{#1#2}}
\newcommand{\flowmodelorigindestination}{\floworigindestination}
\newcommand{\originfunction}[1]{A_{#1}} 
\newcommand{\destinationfunction}[1]{B_{#1}} 
\newcommand{\separationfunction}[2]{S_{#1#2}}

\newcommand{\flow}[1]{\floworigindestination{#1}{}}
\newcommand{\flowmodel}{\flow}
\newcommand{\flowemp}[1]{\empiricalflowsymbol_{#1}}
\newcommand{\destconst}{c}
\newcommand{\sepfunc}[1]{S_{#1}}
\newcommand{\flowproductmodel}[1]{\destconst \originfunction{#1} \sepfunc{#1}}

\newcommand{\distance}[2]{d_{#2}^{(#1)}}
\newcommand{\numsepmeasures}{N_{d}}

\newcommand{\distanceparameter}[1]{\beta^{(#1)}}
\newcommand{\originparameter}{\alpha}
\newcommand{\logdestconst}{\gamma}

\newcommand{\logtraveltimeindex}{1}
\newcommand{\samebundeslandindex}{2}
\newcommand{\samecategoryindex}{3}
\newcommand{\fhindex}{4}
\newcommand{\kirchenindex}{5}
\newcommand{\kunstindex}{6}
\newcommand{\paedindex}{7}
\newcommand{\privatindex}{8}
\newcommand{\uniindex}{9}
\newcommand{\ostenindex}{10}

\newcommand{\logtraveltimesepvar}[1]{\distance{\logtraveltimeindex}{#1}}
\newcommand{\samebundeslanddummy}[1]{\distance{\samebundeslandindex}{#1}}
\newcommand{\samecategorydummy}[1]{\distance{\samecategoryindex}{#1}}
\newcommand{\fhdummy}[1]{\distance{\fhindex}{#1}}
\newcommand{\ostendummy}[1]{\distance{\ostenindex}{#1}}
\newcommand{\kirchendummy}[1]{\distance{\kirchenindex}{#1}}
\newcommand{\kunstdummy}[1]{\distance{\kunstindex}{#1}}
\newcommand{\paeddummy}[1]{\distance{\paedindex}{#1}}
\newcommand{\privatdummy}[1]{\distance{\privatindex}{#1}}
\newcommand{\unidummy}[1]{\distance{\uniindex}{#1}}

\newcommand{\logtraveltimeparameter}{\distanceparameter{\logtraveltimeindex}}
\newcommand{\samebundeslandparameter}{\distanceparameter{\samebundeslandindex}}
\newcommand{\samecategoryparameter}{\distanceparameter{\samecategoryindex}}
\newcommand{\fhparameter}{\distanceparameter{\fhindex}}
\newcommand{\ostenparameter}{\distanceparameter{\ostenindex}}
\newcommand{\kirchenparameter}{\distanceparameter{\kirchenindex}}
\newcommand{\kunstparameter}{\distanceparameter{\kunstindex}}
\newcommand{\paedparameter}{\distanceparameter{\paedindex}}
\newcommand{\privatparameter}{\distanceparameter{\privatindex}}
\newcommand{\uniparameter}{\distanceparameter{\uniindex}}

\newcommand{\dispersionparameter}{\delta}
\newcommand{\inversedispersionparameter}{\dispersionparameter^{-1}}

\newcommand{\significant}{\ensuremath{^*}}
\newcommand{\vsignificant}{\ensuremath{^{**}}}
\newcommand{\esignificant}{\ensuremath{^{***}}}